\documentclass[a4paper]{spie}  

\usepackage[]{graphicx}

\newcommand{\degree}{~$^{\circ}\mbox{C}\;$}

\title{Wide band X-ray Imager (WXI) and Soft Gamma-ray Detector (SGD) for the NeXT Mission} 

\normalsize

\author{T. Takahashi\supit{a,b}, A. Awaki\supit{c}, T. Dotani\supit{a}, Y. Fukazawa\supit{d}, 
K. Hayashida\supit{e}, T. Kamae\supit{f}, 
J.~Kataoka\supit{g}, 
N. Kawai\supit{g}, S. Kitamoto\supit{h},
T. Kohmura\supit{i}, M. Kokubun\supit{b}, K. Koyama\supit{j}, K.~Makishima\supit{b},
H.~Matsumoto\supit{j}, E. Miyata\supit{e}, T. Murakami\supit{k}, 
K. Nakazawa\supit{a}, M. Nomachi\supit{e}, M.~Ozaki\supit{a}, H.~Tajima\supit{f}, 
M.~Tashiro\supit{l}, T. Tamagawa\supit{m}, Y. Terada\supit{m},
H. Tsunemi\supit{e}, T.~Tsuru\supit{j}, K.~Yamaoka\supit{n}, D.~Yonetoku\supit{k}, 
and A.~Yoshida\supit{n}
\skiplinehalf
\supit{a}Institute of Space and Astronautical Science, JAXA, Sagamihara, Kanagawa, 229-8510, Japan; \\
\supit{b}Department of Physics, University of Tokyo, Bunkyo-ku, Tokyo, 113-0033, Japan;\\
\supit{c}Department of Physics, Ehime University, Matsuyama, Ehime, 790-8577, Japan;\\
\supit{d}Department of Physics, Hiroshima University, Higashi-Hiroshima, Hiroshima, 739-8526, Japan;\\
\supit{e}Department of Physics, Osaka University, Toyonaka, Osaka, 560-0043, Japan;\\
\supit{f}Stanford Linear Accelerator Center, Stanford, CA 94309-4349, USA;\\
\supit{g}Department of Physics, Tokyo Institute of Technology, Meguro-ku, Tokyo, 152-8551, Japan;\\
\supit{h}Department of Physics, Rikkyo University, Toshima-ku, Tokyo, 171-8501, Japan;\\
\supit{i}Department of Physics, Kogakuin University, Hachioji, Tokyo, 192-0015, Japan;\\
\supit{j}Department of Physics, Kyoto University, Sakyo-ku, Kyoto, 606-8502, Japan;\\
\supit{k}Department of Physics, Kanazawa University, Kanazawa, Ishikawa, 920-1192, Japan;\\
\supit{l}Department of Physics, Saitama University, Saitama, 338-8570, Japan;\\
\supit{m}RIKEN, Wako, Saitama, 351-0198, Japan;\\
\supit{n}Department of Physics, Aoyama Gakuin University, Sagamihara, Kanagawa, 229-8551, Japan\\
}

\authorinfo{Further author information: (Send correspondence to T. Takahashi)  E-mail: takahasi@astro.isas.jaxa.jp}

\begin{document} 
  \maketitle 

\begin{abstract}
The
 NeXT  mission has been proposed to study high-energy
non-thermal phenomena in the universe. 
The high-energy response 
of the super mirror  will enable us to perform the
first sensitive  imaging observations  up to 80 keV. 
The focal plane detector, which combines 
  a fully depleted X-ray CCD and
   a pixellated CdTe  detector, will provide spectra and images in the wide
   energy range from 0.5 keV to 80 keV.
In the soft gamma-ray band upto $\sim$ 1 MeV, 
a   narrow field-of-view Compton gamma-ray telescope 
utilizing several tens of 
layers of thin Si or CdTe detector will provide  precise spectra
   with much higher sensitivity than present instruments. The continuum
   sensitivity will reach several $\times$ 10$^{-8}$ photons/s/keV/cm$^{2}$ in 
the   hard X-ray region and a few $\times$ 10$^{-7}$ photons/s/keV/cm$^{2}$ in the
   soft $\gamma$-ray region. 
\end{abstract}




\section{Introduction}

The hard X-ray  and gamma-ray bands  have long been recognized as  important windows for
exploring the energetic universe. It is in these energy bands that 
non-thermal emission, primalily due to accelerated high energy particles,
becomes dominant. However, by comparison with the soft X-ray band,
where the spectacular data from 
the XMM-Newton and Chandra satellites are revolutionizing our understanding of
the high-energy universe, the sensitivities of hard X-ray missions flown so far, or currently
under construction, have not dramatically improved
over the last decade. 
In order to study the energy content of non-thermal emission and to draw a more complete
picture of the non-thermal universe, observations by highly sensitive
missions are important.

\begin{figure}
\centerline{\includegraphics[height=7cm]{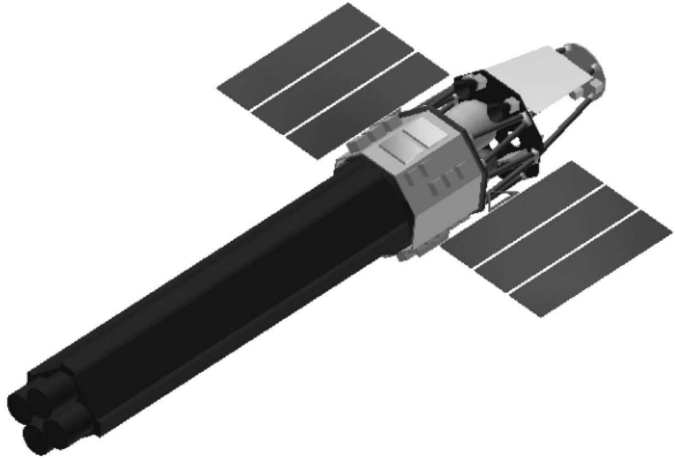}}
\caption{Artist's drawing of the NeXT satellite. The focal  length of the HXT is 12 m. }
\label{Fig:NeXT}
\end{figure}

The NeXT (Non-thermal Energy eXploration
Telescope) mission proposed in Japan is a successor to the Astro-E2 mission,
and is optimised to study the high-energy non-thermal 
universe (Fig.\ref{Fig:NeXT}) \cite{Ref:Proposal,Ref:Kunieda}. 
 NeXT will carry three
hard X-ray telescopes (HXTs) for the wide band X-ray imager (WXI), one 
soft X-ray telescope (SXT)  for the soft X-ray spectrometer (SXS), and a soft
$\gamma$-ray detector (SGD). 
With the first imaging observations up to 80\,keV, 
NeXT is expected to achieve two orders of magnitude improvement
in the sensitivity in the hard X-ray region. The SGD utilizes a new concept of a narrow-FOV 
Compton
telescope\cite{Ref:Takahashi_Yokohama,Ref:Takahashi_SPIE} to improve the sensitivity in the soft $\gamma$-ray region for matching
with the sensitivity of the HXT/WXI combination.
 The extremely low background brought by the SGD will allow
us to measure the precise $\gamma$-ray spectrum upto 1 MeV.
Furthermore, as described in the separate paper\cite{Ref:Tajima-pol}, the SGD will
be  sensitive to the polarization of the  incident $\gamma$-rays, and will
 potentially be the first hard X-ray polarimeter in orbit.
In addition to these high energy instruments, the soft X-ray spectrometer (SXS) will map out the
velocity field in objects  by means of X-ray line
profiles. The SXS is based on a transition edge sensor (TES) and has
both spectral and spatial resolution much better than the X-ray
calorimeter (XRS) onboard the Astro-E2\cite{Ref:Mitsuda}.
In this paper, we describe  the current ideas about the WXI and the SGD.

\section{Wide-band X-ray Imager (WXI) }

The non-imaging instruments flown so far were essentially limited to studies of sources with 10-100 keV fluxes of at best $> $10$^{-12}$ $-$ 10$^{-11}$ erg cm$^{-2}$s$^{-1}$. This limitation is due to the presence of high un-rejected backgrounds from particle events and Cosmic X-ray radiation, which increasingly dominate above 10 keV. Imaging, and especially focusing instruments have two tremendous advantages: first, the volume of the focal plane detector can be made much smaller than for non-focusing instruments, and second, the residual background,  often time-variable, can be measured simultaneously with the source, and can be reliably subtracted. 



\begin{figure}
\centerline{\includegraphics[height=10cm,angle=-90]{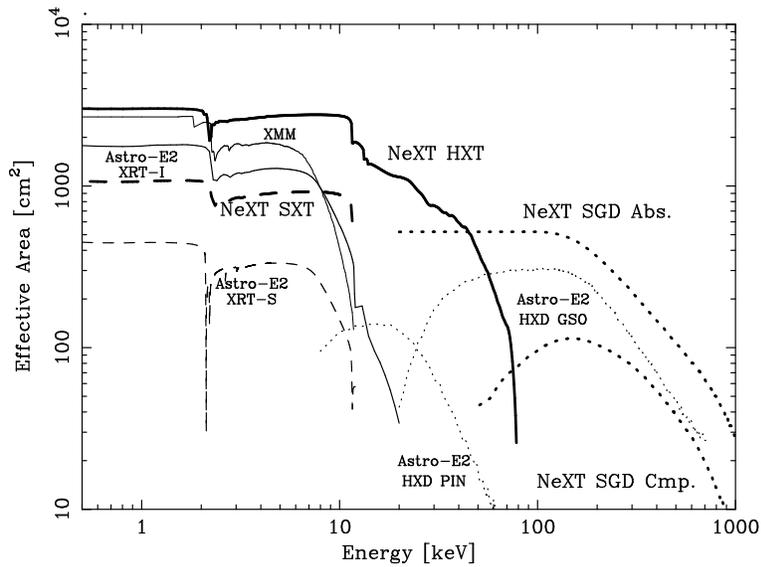}}
\caption{Effective area of the HXT and the SXT as compared to XMM/Newton and Astro-E2. Effective
area of the gamma-ray detectors are also shown for the HXD/Astro-E2 and the SGD/NeXT.  }
\label{Fig:NeXT-Area}
\end{figure}

A depth-graded multi-layer mirror, referred to as a {\it super
mirror}, reflects X-rays not only by total external reflection but
also by Bragg reflection.  A super mirror consists of a stack of
multi-layers with different sets of periodic length and number of
layer pairs. The current mirror design of the HXT
for NeXT\cite{Ref:Furuzawa,Ref:Ogasaka} is based on a carbon/platinum coating, which is adopted in
the InFOC$\mu$s project by Nagoya University and NASA's GSFC\cite{Ref:Ogasaka_1}. 
With three HXT
units, effective areas of 1100 cm$^{2}$ at 20 keV and 230 cm$^{2}$ at
60 keV can be achieved with a focal length of 12\,m\cite{Ref:Proposal}.
Fig. \ref{Fig:NeXT-Area} compare the effective area of the three HXTs with XMM-Newton and
Astro-E2. The effective area of the Soft X-ray Telescope (SXT)  and the SGD
are also shown in the figure.

\begin{figure}
\centerline{\includegraphics*[width=12cm]{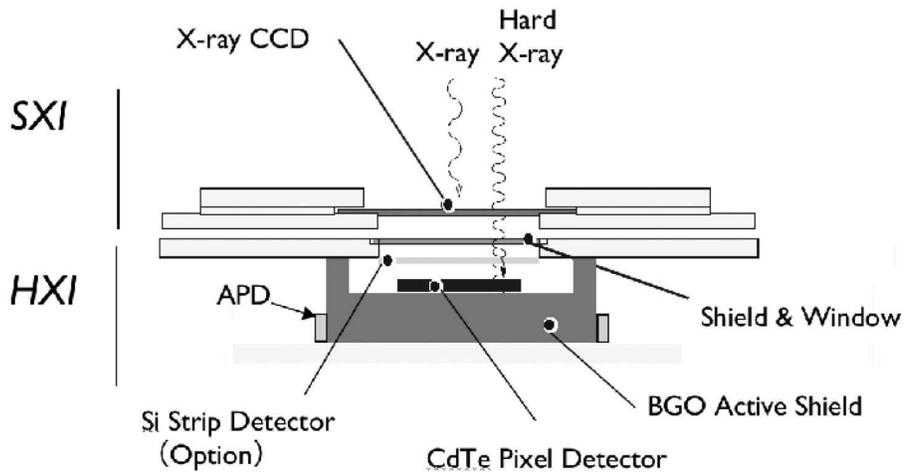}}

\caption{Conceptual drawing of the WXI
proposed for the NeXT mission. The WXI is composed of two sub-instruments: the soft X-ray
Imager (SXI) and the hard X-ray Imager (HXI). The CdTe pixel detector in the HXI is
placed beneath the X-ray CCD in the SXI.  In the WXI, soft X-rays are
 absorbed in the CCD, while hard X-rays penetrate through the CCD and are absorbed 
 in the CdTe pixel detector. The Si strip detector is considered as an
option to shield fluorescence lines from CdTe.}
\label{Fig:WXI_Concept}
\end{figure}

In order to match the energy range covered by the super mirror
(0.5 $-$ 80\,keV), the focal plane detector is required to cover a very
wide energy band. To this end, we have proposed a new focal plane detector
based on an idea of combining  a fully depleted CCD  and a CdTe (Cadmium Telluride)
pixel detector\cite{Ref:Takahashi_NIM,Ref:Takahashi_Yokohama} as the  WXI \cite{Ref:Takahashi_NeXT,Ref:Tsuru_NeXT}. 
 Figure~\ref{Fig:WXI_Concept} is  a schematic design of the WXI showing the concept. 
 The WXI is composed of two sub-instruments; the soft X-ray
Imager (SXI) and the hard X-ray Imager (HXI). Specifications of the SXI and the HXI are
listed in Table~\ref{Table:WXI}. The SXI is based on an
X-ray CCD with very thin dead layer in the device 
and the HXI is based on a fine-pitch CdTe pixel
detector.  In the WXI, soft X-rays will be absorbed in the CCD, while hard X-rays
will penetrate through the CCD and be absorbed in the CdTe pixel detector.
Semiconductor detectors with high mass absorption coefficient, such as CdTe,
are crucial for the
detection of hard X-ray photons.
For CdTe, even a detector with a
thickness of 0.5 mm provides a good detection efficiency for the hard X-ray region
covered by the HXT.
As shown in Fig.~\ref{Fig:WXI_Concept}, the CdTe
detector mounted in the HXI is shielded by a BGO (Bi$_{4}$Ge$_{3}$O$_{12}$) scintillator.
 This shield  is indispensable, as the non X-ray
background is the dominant source of the background in the energy range of the HXI.
A thickness of  $\sim$2 cm for BGO will be required,
 not only for shielding against background photons but
also for reducing the number of particles that reach the CdTe detector
and give rise to activation.

\begin{table}
\caption{Specification for the HXI and the SXI\cite{Ref:Proposal}}

\begin{center}
\begin{tabular} {|lll|}
\hline
HXI & Energy Range & 8 -- 80 keV \\
& Energy Resolution & 0.5 -- 1 keV (FWHM) \\
& Trigger Threshold & $\sim$ 8 keV \\
& Pixel Size & 200 $\sim$ 500 $\mu$m \\
& Detector Size & 20 mm $\times$ (20 -- 30) mm \\
& Timing Resolution & $<$ 100 $\mu$s\\
& Operating Temperature & $-$50 -- 0 degree \\
\hline
SXI & Energy Range & 0.5 -- 20 keV \\
 & Energy Resolution & 130 eV (FWHM at 6keV) \\
 & Pixel Size & 27 $\times$ 27  $\mu$m$^{2}$\\
& Detector Size & 42 mm $\times$ 42 mm \\
& Operating Temperature & $-$90 degree \\
\hline

\end{tabular}
\end{center}
\label{Table:WXI}
\end{table}

The SXI  detects  soft X-rays below 10-20 keV with the high position resolution 
of 27 $\mu$m.  It will be
 based on the technology which has been accumulated
for the Japanese MAXI mission\cite{Ref:Miyata}. 
The SXI utilizes a single, large-format CCD which
covers 12$\times$12 arcmin$^2$ at the focal plane.
An energy resolution of 130 eV (FWHM) at 6 keV is expected at the operating
temperature of $-$90  deg.
SXI utilizes a thinned CCD, whose undepleted layer
is largely removed.
This is necessary to make it transparent to hard X-rays.
If we adopt a backside-illuminated CCD, no extra process is
necessary as the undepleted layer is already removed.
The CCD support and cooling structures are configured near the rim 
of the CCD so as not to interfere with the transmission  of hard X-rays.
Thus the transparent portion to hard X-rays is smaller
than the size of the CCD itself. In order  further to reduce the undepleted layer and 
 to improve the high energy
response, the development of fully depleted  p-channel CCD fabricated on n-type high-resistivity silicon is now under way\cite{Ref:Takagi}.

The current goal for the CdTe detector in the HXI is a pixel detector with both a fine
position resolution of 200$-$250 $\mu$m and a high energy resolution 
better than 1 keV (FWHM), in the energy range from 5 keV to 80 keV. 
Since significant progress has been made recently on the development of 
CdTe technologies\cite{Ref:IEEE2000-1}, it is now possible to fabricate a single crystal CdTe
device with the size of 2$\times$2 cm$^{2}$.
Signals from the individual
pixel electrodes formed on the surface of the CdTe wafer are fed into the readout circuit
built in the ASIC. To
realize the fine pitch CdTe  pixel detector for the HXI, a low noise 
front-end ASIC  with more than several thousand independent
channels will be the key technology. Development of a simple and robust connection
technology is also necessary, because high compression and/or
high ambient temperature would damage the CdTe. 
 In order to cooperate with the BGO shield, the detector should
have fast timing resolution of 10 to 100 $\mu$s, such that we can veto the events
when there is a hit in the shield. 
Taking these requirements into consideration, we have been working on
 high performance
CdTe detectors for both planar and pixel configuration\cite{Ref:Takahashi_IEEE2002,Ref:Nakazawa}.  
Based on  the CdTe wafers manufactured 
from the single crystal grown by the 
Traveling Heater Method (THM-CdTe) and 
with the technology for gold-stud bump bonding, we have developed
CdTe detectors with high energy resolution\cite{Ref:Tanaka,Ref:Oonuki}.

\begin{figure}
\centerline{\includegraphics[width=10cm,clip]{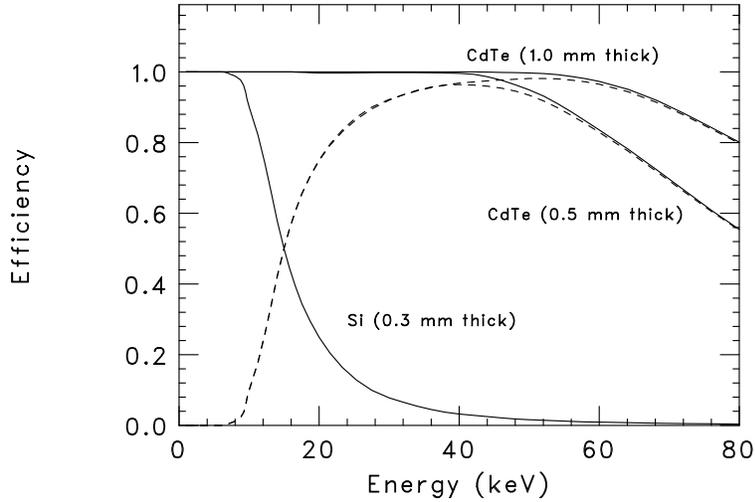}}
\caption{Efficiency of the SXI with 300 $\mu$m thick Si CCD and the HXI for 0.5 mm and 1.0 mm thick
CdTe detector. Effects of a possible dead layer in the Si CCD are not included in the calculation.}
\label{Fig:WXI_Area}
\end{figure}

With NeXT,  we expect to achieve  an area of about 750\,cm$^{2}$ at 30\,keV with a typical angular
resolution of 30$"$. Fig.~\ref{Fig:WXI_Area} shows the efficiency of the WXI with
a 300 $\mu$m thick X-ray CCD and a 0.5 or 1.0 mm CdTe pixel detector. In the calculation, effects of 
a possible dead layer in  the
CCD are not taken into account. By assuming a background level of 1$\times$10$^{-4}$ counts/s/cm$^{2}$/keV,  in which a non X-ray background
is dominant, the source detection limit in 100\,ksec
 would be roughly
10$^{-14}$\,erg\,cm$^{2}$\,s$^{-1}$ in terms of the 10--80\,keV flux for
a power-law spectrum with a photon index of 2. This is about two
orders of magnitude better than present instrumentation.

\begin{figure}
\centerline{\includegraphics[width=7.5cm,clip]{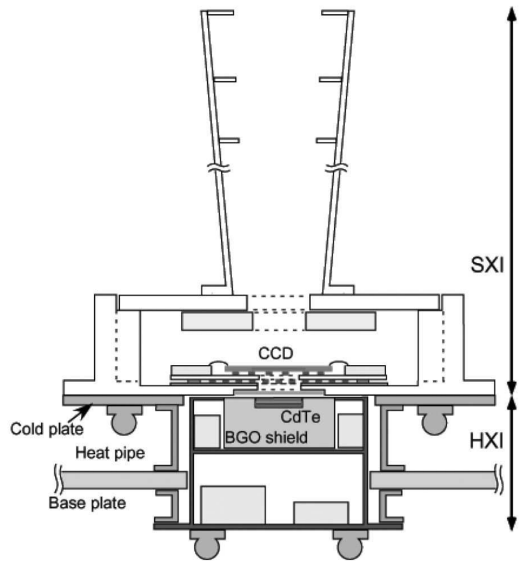}}
\caption{Schematic drawing of the WXI. The SXI and the HXI will be operated at different 
temperatures.}
\label{Fig:WXI_Schematics}
\end{figure}

\section{Narrow FOV Compton telescope}

Highly sensitive observations in the energy range above the HXT/WXI bandpass
is crucial to study the spectrum of   accelerated particles. The SGD 
outperforms previous soft-$\gamma$-ray instruments in  
background rejection capability by adopting a new concept of
narrow-FOV Compton telescope\cite{Ref:Takahashi_NeXT,Ref:Takahashi_Yokohama,Ref:Takahashi_SPIE}. 
 In the energy range above one hundred keV (sub-MeV region),
shielding against background
photons becomes important yet difficult.
The  well-type phoswich counter, which we developed for the WELCOME balloon experiment\cite{Ref:Kamae-Balloon,Ref:Takahashi-Balloon}
and also for  the Astro-E Hard X-ray Detector
 (HXD)\cite{Ref:Kamae,Ref:Tashiro,Ref:Makishima,Ref:Kokubun,Ref:Kawaharada} 
 is a possible solution to achieve a
very low background rate. The phoswich configuration and a tight and active ``well-type'' shield made of  BGO   scintillators is expected to 
reduce  the background to $<$10$^{-4}$ counts/cm$^{2}$/s/keV, 
almost the limit achieved by the configuration
of active shield and collimator.

\begin{table}
\caption{Specification of the SGD\cite{Ref:Proposal}}
\begin{center}
\begin{tabular} {|lll|}
\hline
SGD& Energy Range & 10 keV -- 1 MeV \\
& Energy Resolution & 2 keV (FWHM, 40 keV) \\
& Geometrical Area & 625 cm$^{2}$ \\
& Effective Area & 525 cm$^{2}$ (100 keV) \\
& & 110  cm$^{2}$ (500 keV) \\
& & 100  cm$^{2}$ (Compton mode, 100 keV) \\
& & 40  cm$^{2}$ (Compton mode, 500 keV) \\
& Opening Angle & 0.6 $\times$ 0.6 deg$^{2}$ ($<$ 100 keV) \\
&  & 4 $\times$ 4 deg$^{2}$ ($>$ 100 keV) \\
& (Angular Resolution) & 1.5 deg (RMS, Compton mode, 500 keV)\\
& Operating Temperature & $-$20 -- 0 degree \\
\hline

\end{tabular}

\end{center}
\label{Table:SGD}
\end{table}

\begin{figure}
\centerline{\includegraphics*[width=11.5cm]{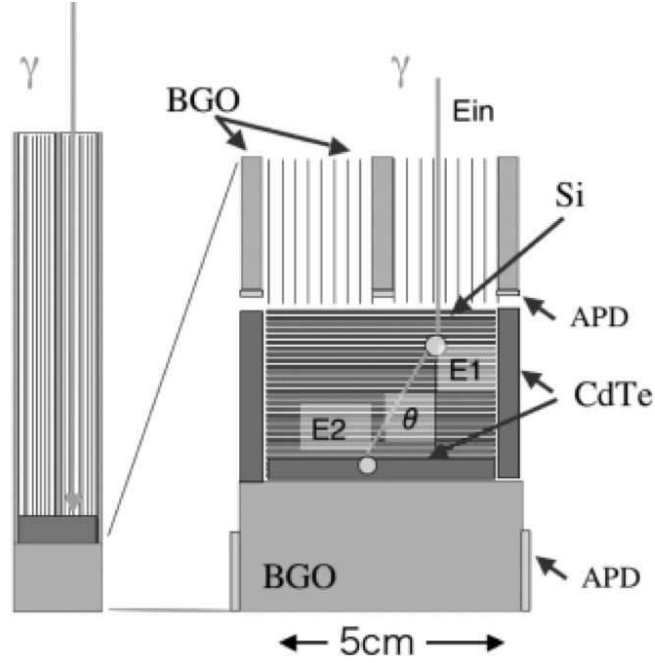}}
\caption{Conceptual design of the SGD module. A stack of 24 Si DSSDs and  CdTe pixel detectors are assembled to form a semiconductor Compton Telescope. 
 The telescope  is mounted inside the bottom of a
well-type active shield. The opening angle provided by the BGO shield is 
 4 deg at 500 keV.  An additional 
 copper collimator restricts the field of view of the telescope to 30' for low energy photons ($<$100 keV).
 With this narrow FOV, events are rejected as background, if the reconstructed Compton ring does not intercept the FOV. }
 \label{Fig:SGD_Concept}
\end{figure}

\begin{figure}
\centerline{\includegraphics*[width=13.2cm]{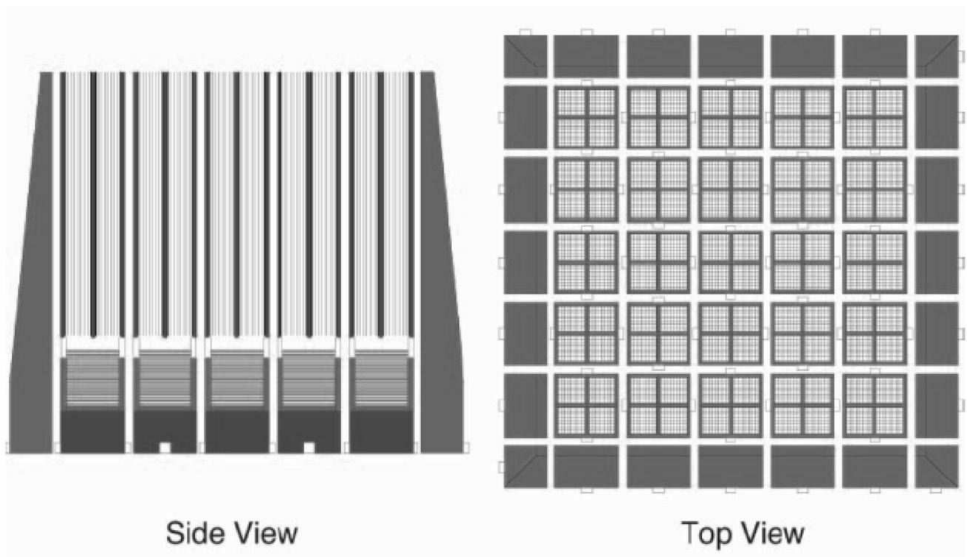}}
\caption{A schematic drawing of the SGD. It consists of  25 units of the narrow FOV 
Compton telescope surrounded by 24 shield counters made of BGO. The geometrical area of 525 cm$^{2}$ is currently assumed for
the sum of 25 units. If we select an event that has two hits in the detector (and no hit in the
BGO shields) and if we require the proper reconstruction for the Compton kinematics, 
the effective area at 200 keV becomes $\sim$ 100 cm$^{2}$, including the reconstruction 
efficiecy.}
\label{Fig:SGD_ALL}
\end{figure}

\begin{figure}
\centerline{\includegraphics*[height=7.5cm]{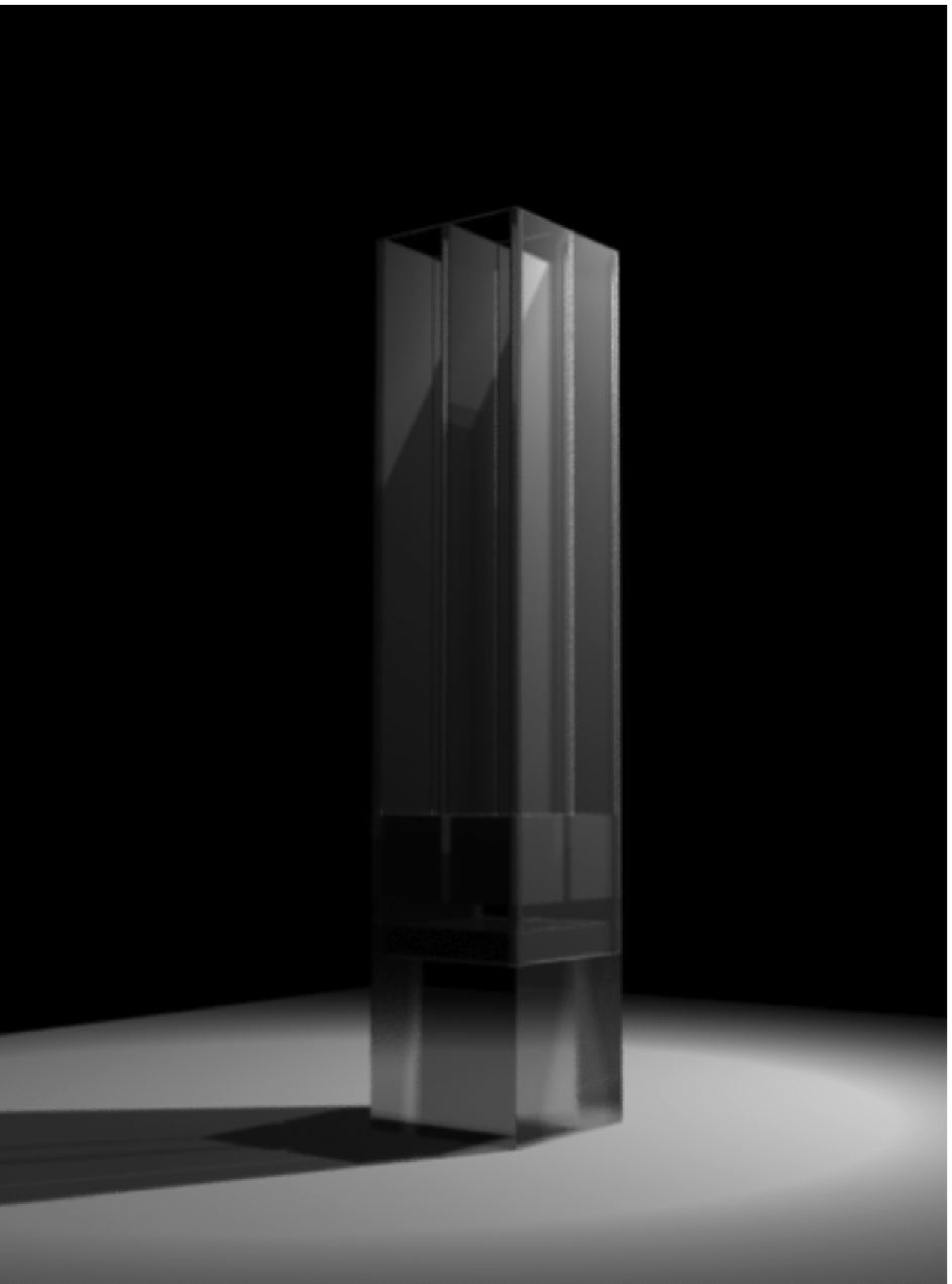}\hspace{1cm}
\includegraphics*[height=7.5cm]{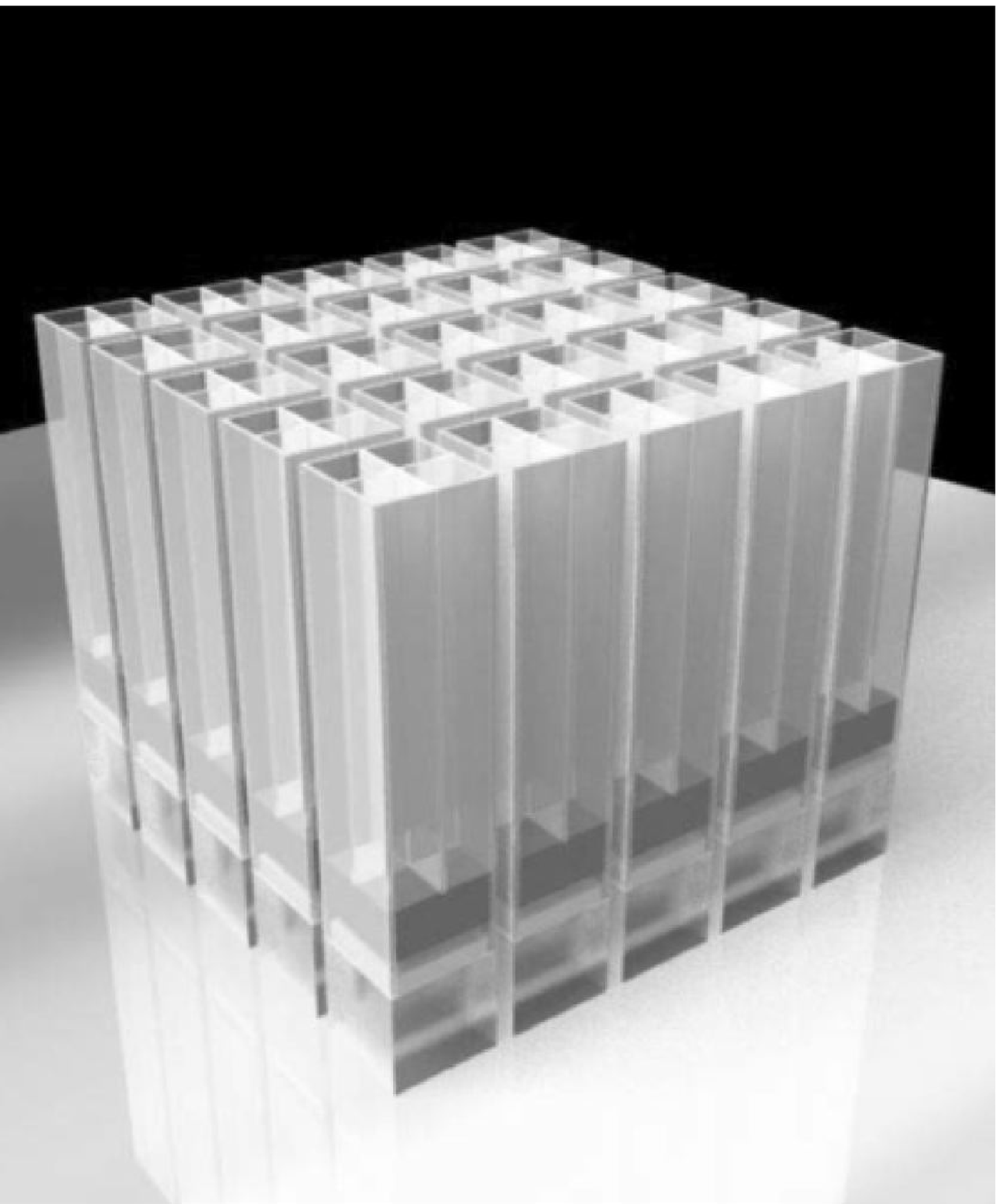}}
\caption{Artist drawing of the SGD for one unit (left) and 25 units (right). The side BGO shield
is not shown. }
\label{Fig:SGD_Computer}
\end{figure}

In order to  lower the background dramatically and thus to improve the sensitivity
as compared to the HXD,  we combine a
stack of Si strip detectors and CdTe pixel detectors to form a
Compton telescope. 
The telescope  is then mounted inside the bottom of a
well-type active shield. As shown  schematically  in Fig.\ref{Fig:SGD_Concept}, 
 the telescope consists 
 of 24 layers of DSSDs (double-sided silicon strip detectors) and 2 layers of thin CdTe pixellated detectors surrounded by 5 mm thick CdTe pixellated detectors. Specifications for the SGD and for each component
 for the Compton telescope are listed in Table \ref{Table:SGD} and Table \ref{Table:SGD_Ele}, respectively.
The opening angle provided by the BGO shield is 
 4 degree at 500 keV. As compared to the HXD, the  shield part is made compact
 by adopting the newly developed avalanche photodiode\cite{Ref:Kataoka,Ref:Nakamoto}.
An additional 
 copper collimator restricts the field of view of the telescope to 30' for low energy photons ($<$100 keV)
 to minimize the flux due to the Cosmic X-ray Background from the FOV.  
 These modules are then arrayed to provide the required area (Figure \ref{Fig:SGD_ALL}). Figure \ref{Fig:SGD_Computer} shows a drawing of the design goal of the SGD instrument which consists of a 5 $\times$ 5 array of identical detector modules. 
 
\begin{table}
\begin{center}
\caption{Specification of the DSSD and the CdTe pixel detector for the SGD\cite{Ref:Proposal}}

\begin{tabular}{|lll|}
\hline
DSSD & strip pitch & 0.4 mm\\
 & thickness & 0.5 mm\\
 & number of strips & 125 \\
 & active area & 5 $\times$ 5 cm$^{2}$ \\
 & energy resolution & 1.5 keV (FWHM) \\
\hline
CdTe pixel detectors & pixel size & 2 $\times$ 2 mm$^{2}$ \\
& thickness & 0.5 mm (thin), 5 mm (thick) \\
& number of pixels & 25 $\times$ 25 \\
& active area  & 5$\times$ 5 cm$^{2}$ \\
& energy resolution & 1.5 keV (FWHM) \\
\hline
\end{tabular}
\end{center}
\label{Table:SGD_Ele}
\end{table}

An important feature of the SGD is that we can
require each event to interact twice in the stacked
detector, once by Compton scattering in a stack of Si strip
detectors, and then by photo-absorption in the CdTe part (Compton
mode).
Once the locations and energies of the two interactions are
measured, the Compton kinematics allows us to calculate the energy and
direction (as a cone in the sky) of the incident $\gamma$-ray  by following the Compton equation,
 \begin{equation}
\label{Eq:Compton}
 E_{in} = E_{1} + E_{2},
 \end{equation}
 \begin{equation}
\cos \theta = 1 - m_{e}c^{2}  (\frac{1}{E_{2}} -\frac{1}{E_{1}+E_{2}} ),
\end{equation}
where E$_{1}$ is the energy of the recoil electron, E$_{2}$ is the energy of the scattered
photon and $\theta$ is the scattering angle. 
The high spectral and spatial resolution of Si and CdTe semiconductor detectors
allow the instrument to achieve  high angular resolution. 
Since the major interaction of photon above 60 keV in Si is Compton scattering, the stack of DSSDs acts
as an efficient scatterer especially for the low energy region below 300 keV.  
Regarding   the uncertainty of the  order of scattering in an event,
 we can use the relation that
 the energy deposition by Compton scattering
is always smaller than that of the photo-absorption for energies 
below $E_{\gamma}$ = 255 keV
($E_{\gamma} = m_{e}$/2). This relation holds above this energy,
 if the scattering angle $\theta$ is smaller than 
$\cos^{-1} (1 - \frac{1}{2}(\frac{m_{e}}{E_{\gamma}})^2)$\cite{Ref:Takahashi_NeXT}.  
  
  The major advantage of employing a narrow FOV 
is that the direction of incident $\gamma$-rays is
constrained inside the FOV. 
 If the Compton ring does not intercept the FOV, we can reject the event
as background. 
Most backgrounds can be rejected by  requiring 
this condition
(albeit with an corresponding reduction in instrument
effective area).
Background photons from the BGO and copper collimator 
for which the reconstructed Compton ring intersects the FOV,  
cannot  be eliminated if there is no signal detected in the active shield,
however this source of background contributes only within a
 limited range of scattering angle. 
Combining background suppression techniques available in the SGD, 
we expect to achieve background levels of 5$\times$10$^{-7}$ counts/s/cm$^{2}$/keV
at $\sim$ 100 keV and   2$\times$10$^{-7}$ counts/s/cm$^{2}$/keV
at $\sim$ 500 keV. As shown in Fig. \ref{Fig:NeXT-Area}, the effective area of the SGD
in the Compton mode is 120 cm$^{2}$ at 200 keV and 50
cm$^{2}$ at 400 keV, if we use 25 units (total geometrical area is 625 cm$^{2}$).

\begin{figure}
\centerline{\includegraphics*[height=6.5cm, width=9cm]{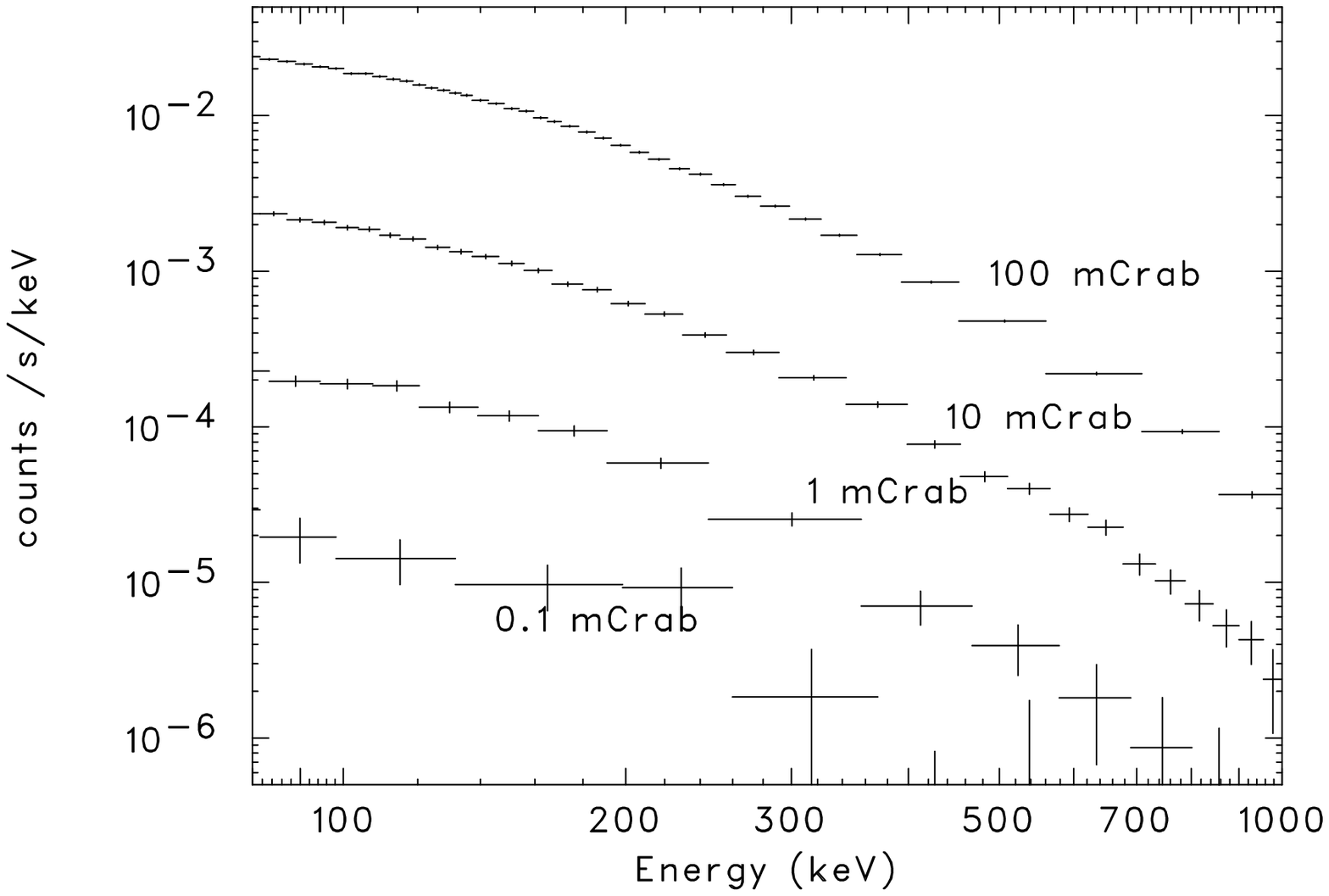}\hspace{5mm}
\includegraphics*[height=6.5cm]{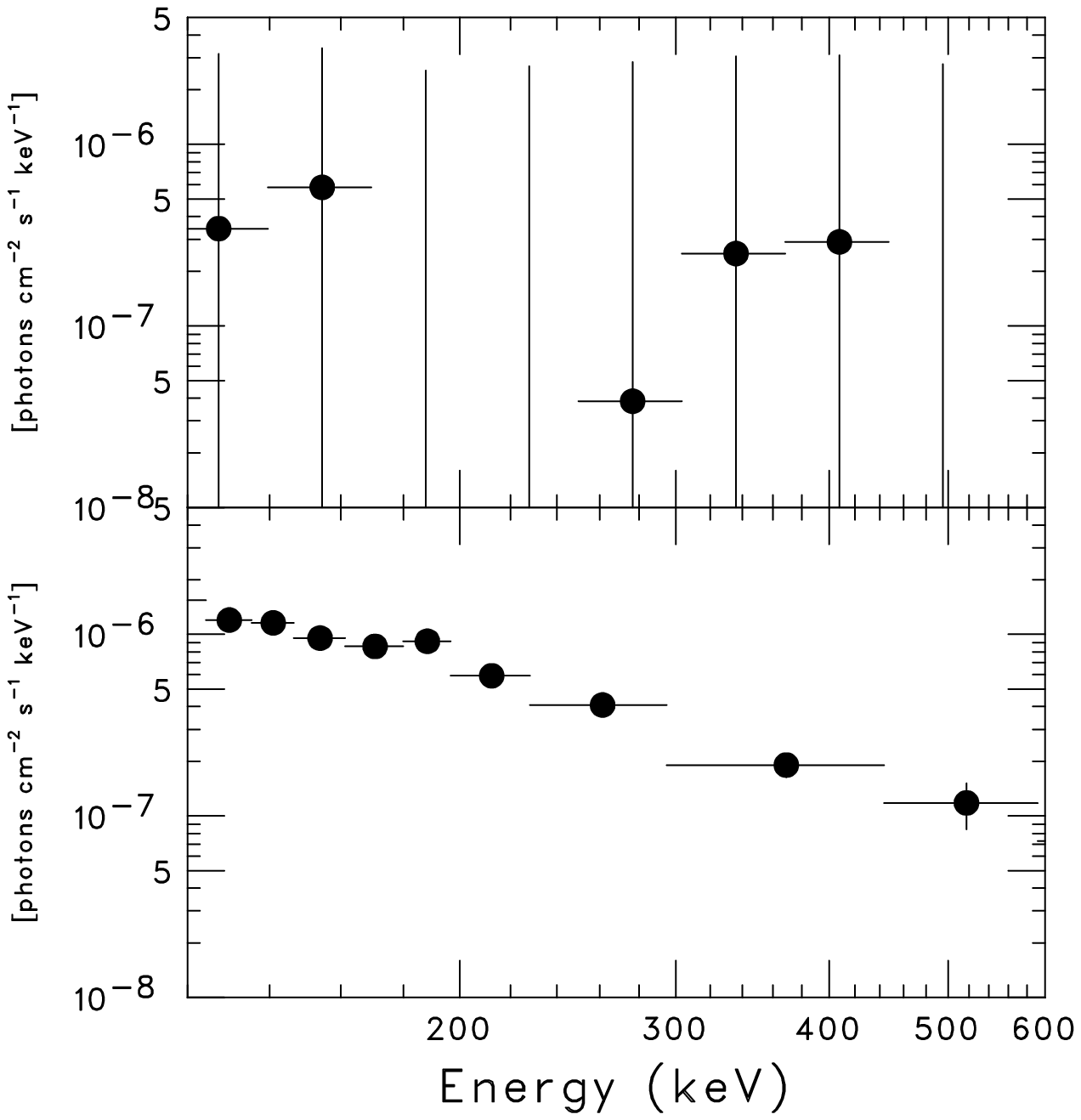}
}
\caption{(left) Expected energy spectrum for a 100 ks observation of 0.1 $-$ 100 mCrab sources with the Compton mode. The background level of 5 $\times$ 10$^{-7}$ counts/s/cm$^{2}$/keV is assumed. (right) Comparison of the energy spectrum for a 100 ks observation of 1 mCrab source expected from the SGD operated in the Compton mode (bottom plot) and the instrument with the effective area of 3300 cm$^{2}$ and the background level of 5  $\times$  10$^{-4}$ counts/s/cm$^{2}$/keV (top plot). A systematic error of 5 \% is included in the background estimation for both cases.}
\label{Fig:Compare}
\end{figure}

The concept of a narrow FOV Compton telescope is expected to 
reduce drastically the background from radio-activation of the detector
materials, which is a dominant background in the case of the
Astro-E2 HXD\cite{Ref:Kokubun}. 
   Figure ~\ref{Fig:Compare}  (left) shows the expected energy spectrum for a 100 ks observation of 0.1 $-$ 100 mCrab sources  by the SGD with Compton mode, under an assumption of  the background level of 5 x 10$^{-7}$ counts/s/cm$^{2}$/keV. Fig.~\ref{Fig:Compare} (right) compares the energy spectrum for a 100 ks observation of 1 mCrab source (photon index 1.7) expected from the SGD Compton mode (bottom plot) and that for an instrument with an effective area of 3300 cm$^{2}$ (50 times the SGD effective area) and a background level of 5 x 10$^{-4}$ counts/s/cm$^{2}$/keV . These
results, derived from simple simulations, show that a high signal-to-background ratio is important
to achieve the high sensitivity.
In addition to the reduction of background, it should be noted that
the 30' FOV of the fine collimator is required to 
improve the sensitivity limited by  source confusion below a few hundred keV. 
With the SGD, we can also measure polarization of
incident gamma-rays from the azimuthal distribution of the Compton
scattered photons\cite{Ref:Tajima-pol,Ref:Mitani}.

\begin{figure}[tbh]
\begin{center}
   \begin{tabular}{ll}
        \includegraphics[height=5.0truecm]{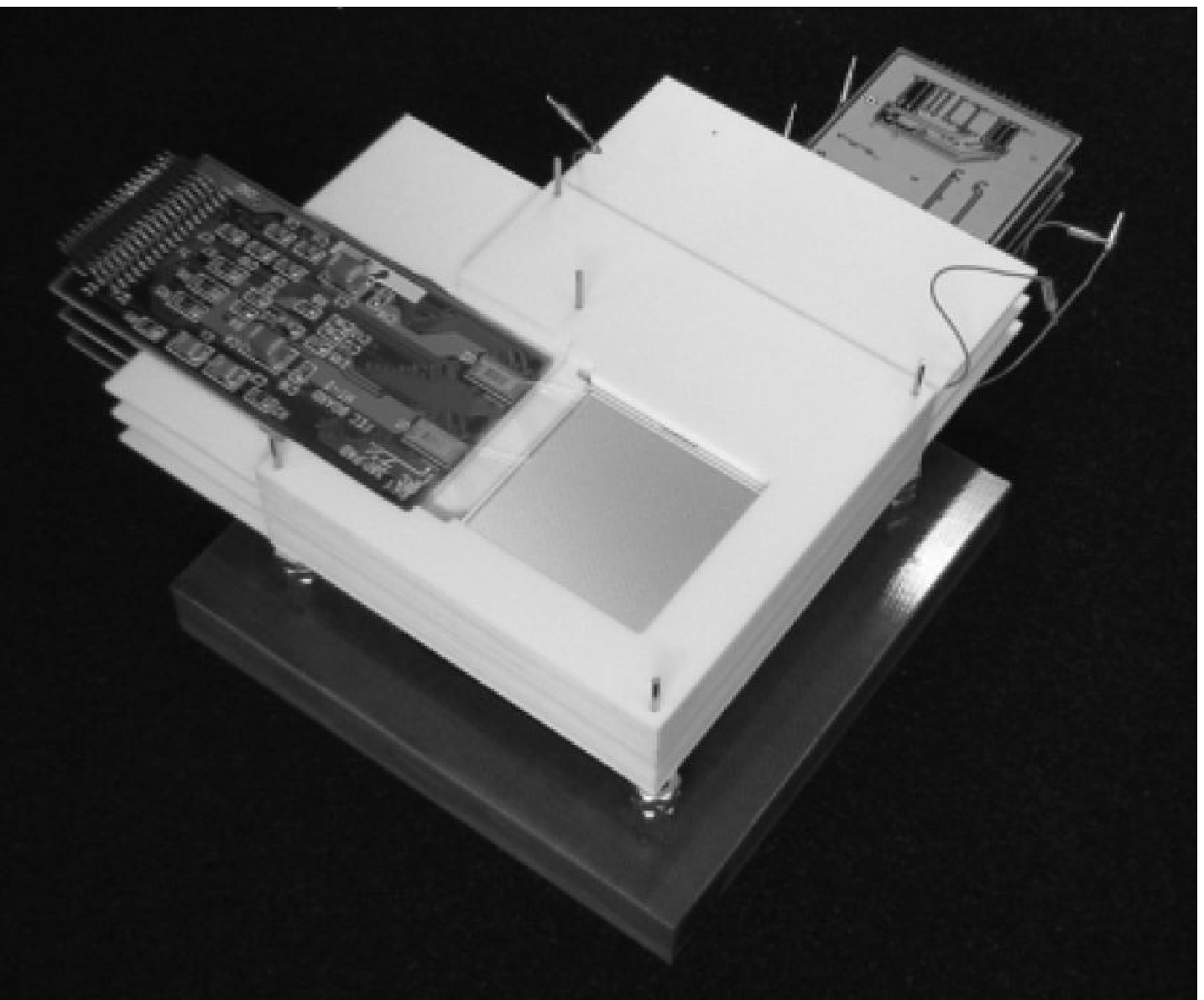}
          & \includegraphics[height=5.0truecm]{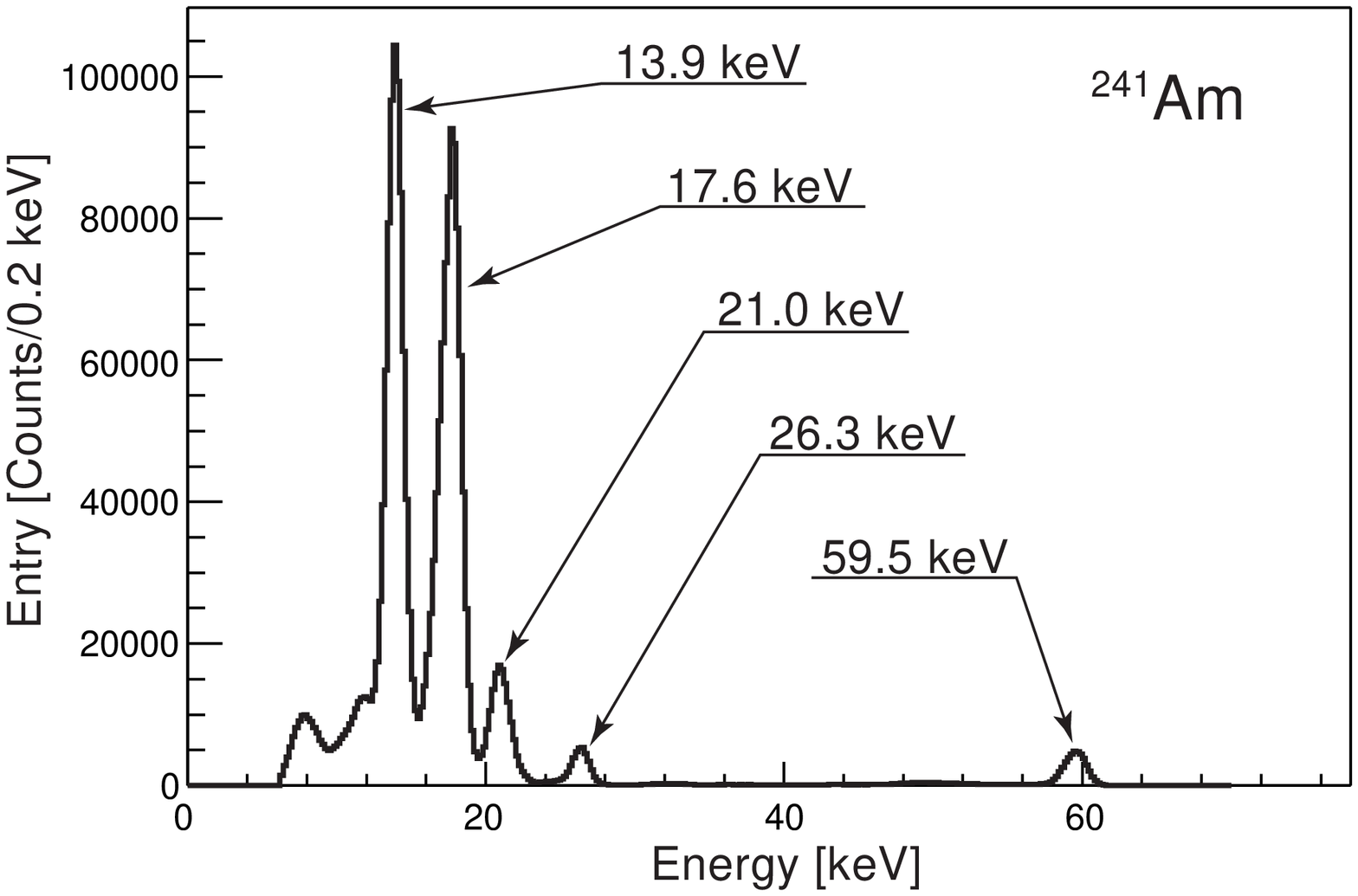}\\
   \end{tabular}
\end{center}
\caption[] {(left) Photo of three layers of DSSD stack developed for the prototype Si/CdTe Compton Camera\cite{Ref:Tajima_DSSD,Ref:Fukazawa} (right) Energy spectrum from DSSD for $^{241}$Am source. The signal from each strip is processed by a newly
developed analog front-end ASIC\cite{Ref:Tajima-ASIC}. An energy resolution of 1.3 keV (FWHM) for 
60 keV can be achieved at 0 \degree.}
\label{Fig:DSSD}
\end{figure}

\begin{figure}[tbh]
\begin{center}
   \begin{tabular}{ll}
        \includegraphics[height=5.0truecm]{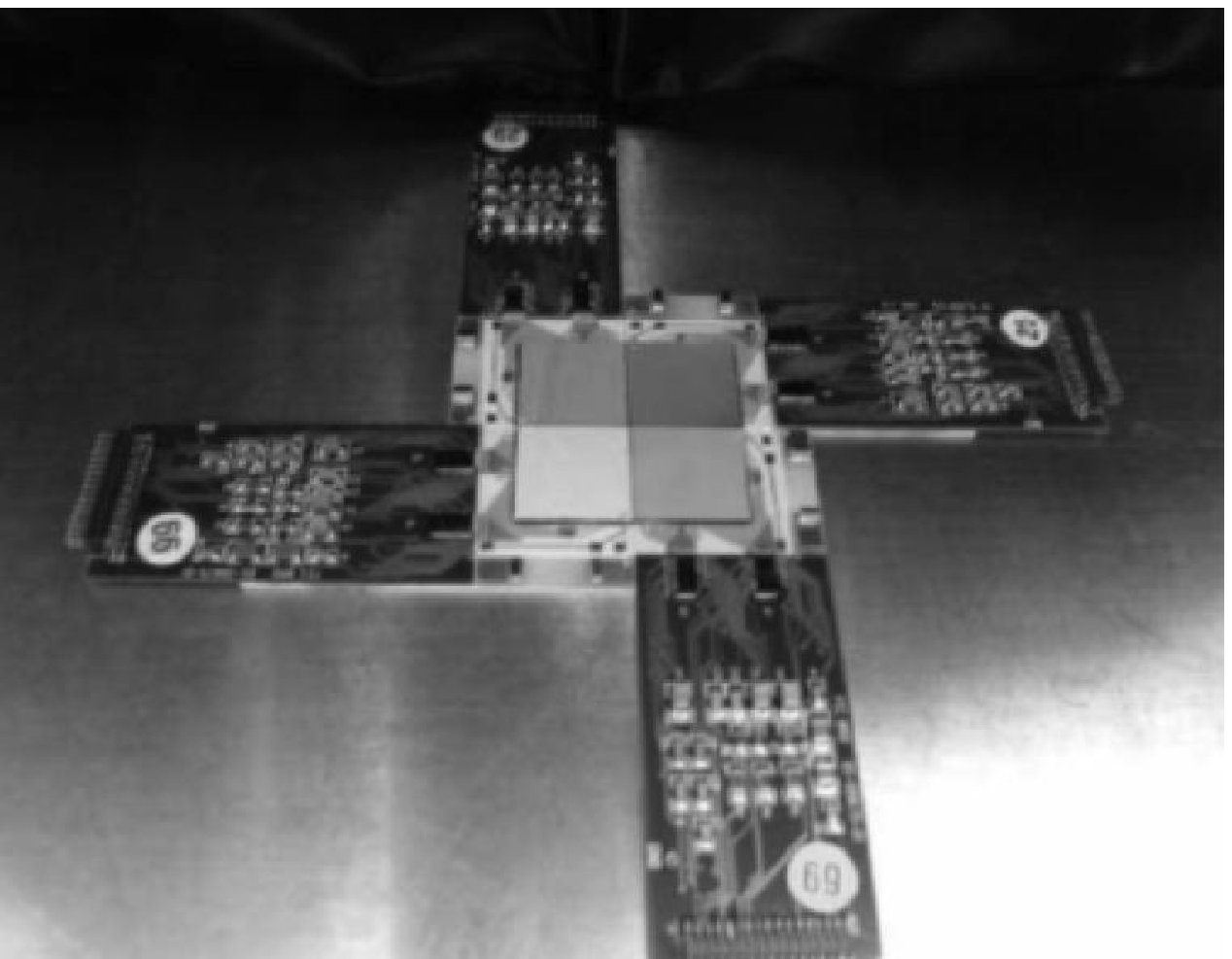}
          & \includegraphics[height=5.0truecm]{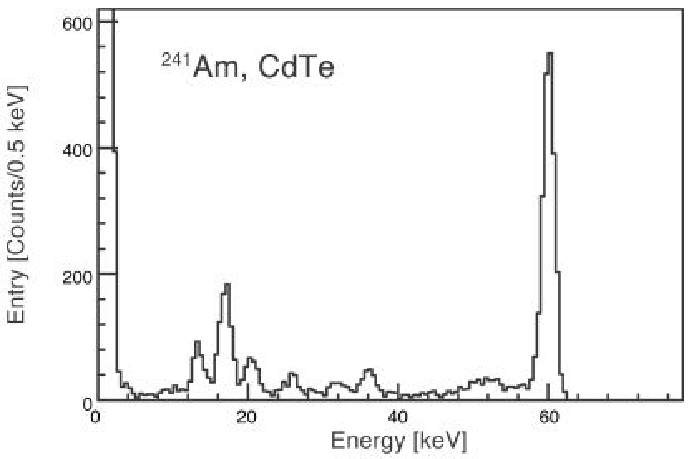}\\
   \end{tabular}
\end{center}
\caption[] {(left) Photo of large area CdTe pixel detectors developed for the prototype Si/CdTe Compton Camera\cite{Ref:Mitani,Ref:Tanaka}. The detector has  an area of 3.2 cm $\times$ 3.2 cm and a thickness of 0.5 mm. Pixel size is 2 mm $\times$ 2 mm.  (right) Energy spectrum from CdTe (one pixel) for $^{241}$Am source. An energy resolution is  1.6 keV (FWHM) for 
60 keV at 0 \degree.}
\label{Fig:CdTe_Pixel2}
\end{figure}

Based on our CdTe and Si detector technologies, we are working on a  Si/CdTe semiconductor
Compton telescope. The photo of DSSD and CdTe pixel detectors used in the prototype and
spectra taken from those detectors are shown in Fig. \ref{Fig:DSSD} and Fig. \ref{Fig:CdTe_Pixel2}.
Performance of the
prototype and  results of polarization measurements are described in other publications\cite{Ref:Tajima_DSSD,Ref:Mitani,Ref:Fukazawa,Ref:Tanaka}.

\section{Summary}

\begin{figure}
\centerline{\includegraphics*[width=8cm,angle=90]{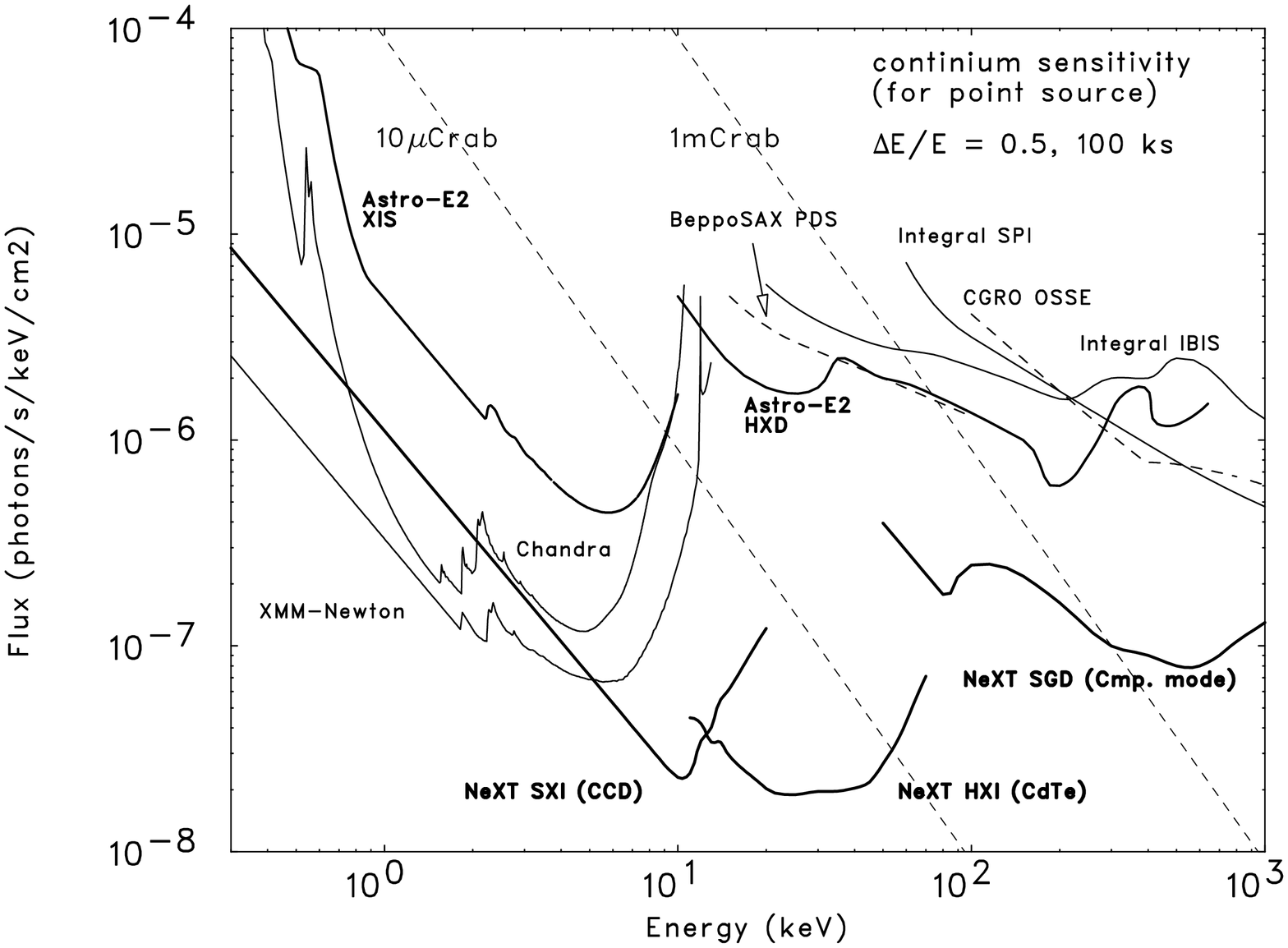}}
\caption{The expected WXI and SGD senstivities for continuum emissions from 
a point source,
assuming an  observation time of 100 ks.}
\label{Fig:Continuum}
\end{figure}

\begin{figure}

\centerline{\includegraphics*[width=8cm,angle=90]{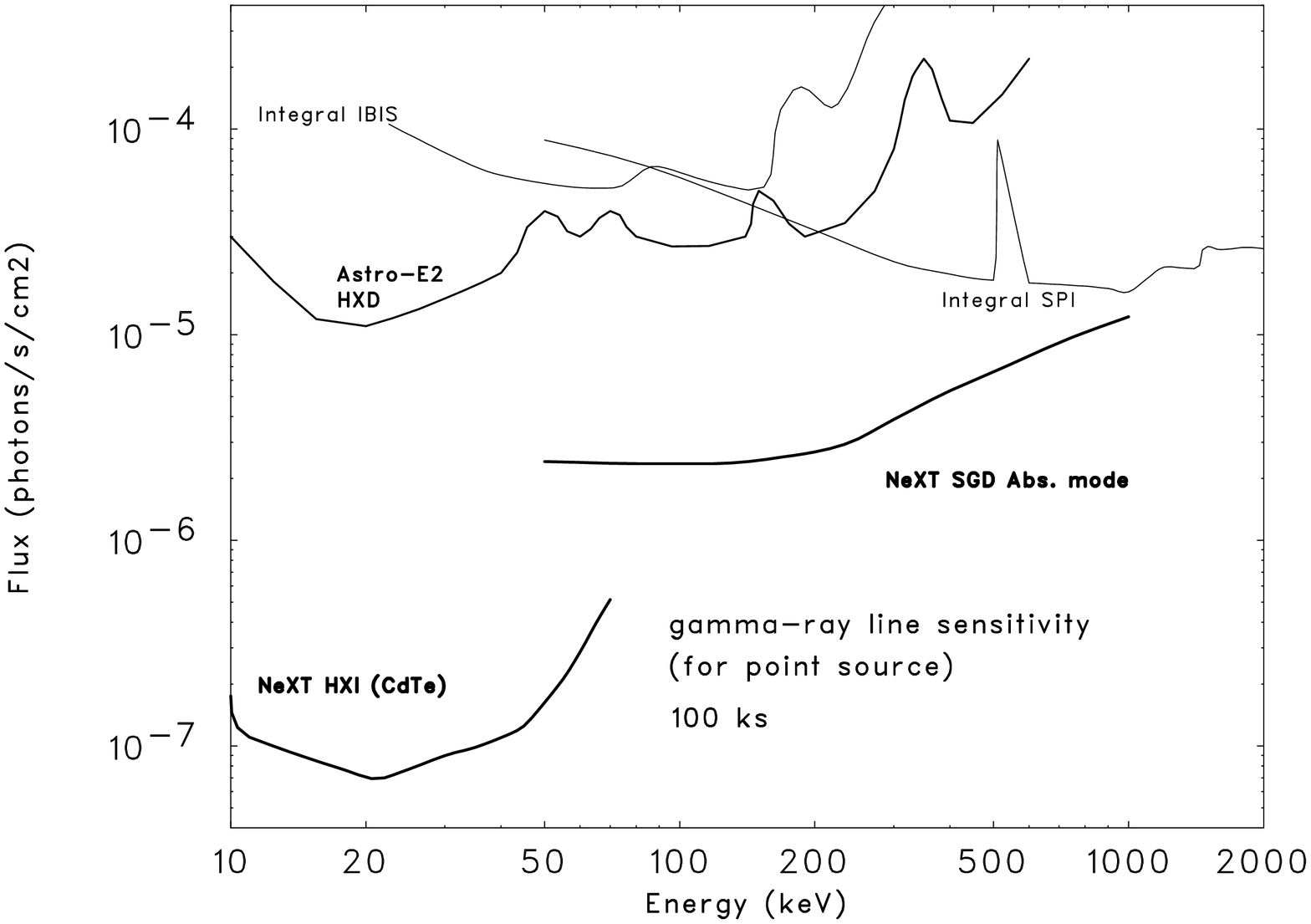}}
\caption{The expected WXI and SGD senstivities for line emissions from a point source,
assuming an  observation
time of 100 ks.}
\label{Fig:Line}
\end{figure}

The line and continuum sensitivities 
of the NeXT mission
 for 100 ks observation are  shown in Fig.\ref{Fig:Continuum} and Fig.\ref{Fig:Line}, together with 
those of  other missions. 
 The continuum
   sensitivity could reach several $\times$ 10$^{-8}$ photons/s/keV/cm$^{2}$ in 
the   hard X-ray region and a few $\times$ 10$^{-7}$ photons/s/keV/cm$^{2}$ in the
   soft $\gamma$-ray region. 
The high-energy response 
of the super mirror  of NeXT will enable us to perform
the first sensitive  imaging observations  up to 80 keV. 
By combining  an X-ray CCD and a  CdTe pixel  detector, the WXI is an ideal solution for the
focal plane detector of the super mirror, providing fine imaging capability and high spectral resolution
with almost full detection efficiency.
The narrow field-of-view Compton $\gamma$-ray telescope realized by the SGD extends the bandpass to well above the cutoff for the hard X-ray telescope and allows us to study the high energy end of the
particle spectrum through the sensitive observation of the  $\gamma$-ray spectrum up to 1 MeV. 
By combining these detectors, we expect to achieve an unprecedented level of sensitivity 
 in the hard X-ray and sub-MeV $\gamma$-ray region for both line and continuum emission.

\end{document}